\documentclass[aps,twocolumn,superscriptaddress,showpacs,amsmath]{revtex4}
\usepackage{graphicx}
\usepackage{bm}
\usepackage{amsmath}
\begin{document}

\title{Spin-Orbital Entanglement and Phase Diagram of
Spin-orbital Chain with $SU(2) \times SU(2)$ Symmetry}

\author{Yan Chen}
\affiliation{Department of Physics and Center of Theoretical and
Computational Physics, The University of Hong Kong, Pokfulam Road,
Hong Kong, China}
\author{Z. D. Wang}
\affiliation{Department of Physics and Center of Theoretical and
Computational Physics,  The University of Hong Kong, Pokfulam Road,
Hong Kong, China} \affiliation{ National Laboratory of Solid State
Microstructures, Nanjing University, Nanjing, China}
\author{Y. Q. Li}
\affiliation{ Department of Physics, Zhejiang University, Hangzhou,
China}
\author{F. C. Zhang}
\affiliation{Department of Physics and Center of Theoretical and
Computational Physics,  The University of Hong Kong, Pokfulam Road,
Hong Kong, China} \affiliation{ Department of Physics, Zhejiang
University, Hangzhou, China}

\begin{abstract}
Spin-orbital entanglement in quantum spin-orbital systems is
quantified by a reduced von Neumann entropy, and is calculated for
the ground state of a coupled spin-orbital chain with $SU(2)\times
SU(2)$ symmetry. By analyzing the discontinuity and local extreme of
the reduced entropy as functions of the model parameters, we deduce
a rich phase diagram to describe the quantum phase transitions in
the model. Our approach provides an efficient and powerful method to
identify phase boundaries in a system with complex correlation
between multiply degrees of freedom.

\end{abstract}


\maketitle Exotic states associated with the orbital degrees of
freedom in transition-metal oxides have attracted considerable
interest recently. Examples of such systems with spin-orbital
couplings include spin-gap materials Na$_{2}$Ti$_{2}$Sb$_{2}$O and
NaV$_{2}$O$_{5}$, manganites La$_{1-x}$Sr$_x$MnO$_3$, and
V$_{2}$O$_{3}$~\cite{Book1,Tokura98,Tokura00,Bao93,Feiner97}.
Intriguing physical properties in these systems include the
emergence of orbital ordering, the appearance of complex coupled
excitations involving both spin and orbital degrees of freedom.
Starting from a multi-band Hubbard model at strong coupling limit
and at the integer fillings of electrons per unit cell, the charge
degree of freedom is frozen and one may derive an effective
spin-orbital model~\cite{Kugel73,Cast78}. One of the simplest such
systems is the $SU(2) \times SU(2)$ model with $SU(2)$ symmetries
for spin-1/2 operator $\mathbf{S}_i$ as well as for pseudospin-1/2
operator $\mathbf{T}_i$ representing two degenerate orbitals. There
have been a lot of activities recently on the one-dimensional
spin-orbital coupled systems ~\cite{gri99,Mila99}, especially on its
phase diagram ~\cite{Pati98,Azaria99,yam,Itoi00,Zheng01}. Rich
quantum phases include both conventional
ferromagnetic/antiferromagnetic gapless phases and symmetry broken
gapped states. In the strong coupling regime where the interplay
between spin and orbital quantum fluctuations is crucial, the
detailed phase diagram still remains controversial and a more
comprehensive understanding is awaited.

More recently, the investigation of quantum
entanglement
from the perspective of quantum information theory has gained much
insight for a deeper understanding of quantum many particle physics,
especially quantum phase transitions. Many theoretical studies have
been devoted to the entanglement in one dimensional spin-1/2
systems~\cite{Wootters98,Wang02,AOsterloh2002,GVidal03,Cirac04,chen04}
and in interacting fermion and boson systems~\cite{Gu04,chen05}.
Quantum entanglement has been quantified in terms of the spin-spin
concurrence~\cite{Wootters98}, contiguous block
entanglement~\cite{GVidal03}, and sublattice
entanglement~\cite{chen04,chen05}. There are evidences to suggest a
close connection between quantum phase transition and local extreme
or singularity of the quantum entanglement when it is measured
appropriately~\cite{chen04}. In the coupled spin-orbital systems,
one expects the spin-orbital entanglement (SOE) to be important. In
particular, near a quantum transition point, one may naturally
expect that the entanglement may manifest itself accordingly. Recent
theoretical study has also demonstrated that the SOE could lead to
the violation of the Goodenough-Kanamori rules~\cite{Khaliullin06}.

In this paper, we propose a reduced von Neumann entropy to quantify
the SOE, which measures the interplay between spin and orbital
degrees of freedom of the quantum states. We use small size exact
numerical method to calculate the reduced entropy of the ground
state of the spin-orbital chain given by Hamiltonian Eq. (1) and
study its relation with the quantum phase transition of the system.
Our results show that this novel measure of the entanglement can
reveal faithfully the quantum transition points and phase boundaries
of the complex phase diagram of the system. Our results indicate
that the strategy of evaluating entanglement measure is powerful and
efficient for extracting valuable information of the quantum
systems.

We consider a one-dimensional spin-orbital Hamiltonian with
$SU(2)\times SU(2)$ symmetry,
\begin{equation}
H=\sum\limits_{i}\left( \mathbf{S}_{i}\cdot\mathbf{S}_{i+1}+x\right)
\left( \mathbf{T}_{i}\cdot\mathbf{T}_{i+1}+y\right) ,
\end{equation}
where $\mathbf{S}_i$ are spin-1/2 operators while $\mathbf{T}_i$
denote the orbital pseudo-spin 1/2 operators. $x$ and $y$ are two
tuning parameters. At $x=y$, the model has an interchange symmetry
between spin and orbital. The model at $x=y=1/4$ is a special case
possessing a higher $SU(4)$ symmetry, and there are three gapless
modes (spin, orbital and spin-orbital) in the low-lying
excitations~\cite{Sutherland75,YQLi98}. It is also known that the
model at $x=y=3/4$ has an exact ground state, in which the spin and
orbital form dimerized singlets in a staggered pattern, and the
doubly degenerate ground states can be expressed as gapped matrix
product state~\cite{Kolezhuk98}.

In the spin-orbital model, the importance of the SOE has long been
recognized~\cite{YQLi98,Khaliullin06}. However, a quantitative
measure for the entanglement is still lacking. We propose to measure
the SOE by a reduced von Neumann entropy defined as
\begin{eqnarray}
S^{so} := - \rm{tr}_s \left( \rho_{s} \log_2 \rho_{s} \right),
\label{ent-ent}
\end{eqnarray}
where $\rho_s\equiv \rm{tr}_{o} |{\Psi}\rangle\langle{\Psi}|$ is the
reduced density matrix of the spin part in the state
$|{\Psi}\rangle$ by integrating out all the orbital degree of freedom.
Obviously, Eq. (2) gives $S^{so}=0$ if spin $\mathbf{S}$ and orbital
$\mathbf{T}$ are decoupled. The motivation for such a measure is to
better reveal the correlation between two distinctive degrees of
freedom. This measure is similar to the recent proposal of the
reduced entropy $S_L$ of a block of subsystem in study of the
relations between entanglement and quantum phase transition, where
$S_L$ is defined by
\begin{eqnarray}
S_L := - \rm{tr} \left( \rho_{L} \log_2 \rho_{L} \right),
\label{ent-ent2}
\end{eqnarray}
where $\rho_L\equiv \rm{tr}_{L} |{\Psi}\rangle\langle{\Psi}|$ is the
reduced density matrix for a block of subsystem $B_L$. The analogy
of the SOE with the block subsystem becomes more clear if we map the
model of Eq. (1) onto a two-leg "spin" ladder system with one chain
described by spin $\mathbf{S}$ and the other chain by orbital
$\mathbf{T}$ and the two sites on each leg are coupled by a
four-operator interactions~\cite{Nersesyan97}.

Let us first examine the SOE defined in Eq. (2) in a few simplest
cases. For a single site system, the SOE has a one-to-one
correspondence to the two pure spin-1/2 system with one spin for
$\mathbf{S}$ and the other for $\mathbf{T}$. For a two-site system,
the spin (orbital) states can be either a singlet $|{\Psi}_S^s
\rangle$ ( $|{\Psi}_O^s \rangle$) or triplet $|{\Psi}_S^{t} \rangle$
( $|{\Psi}_O^{t} \rangle$). It is easy to check that $S^{so}=0$ for
all the spin-orbital decoupled states, and $S^{so}=1$ for the state
$1/\sqrt 2 (|\Psi_S^{s} \rangle|\Psi_O^{t} \rangle \pm |\Psi_S^{t}
\rangle |\Psi_O^{s} \rangle)$. Next we proceed to the 4-site (1234)
cluster, which  is the smallest system size to have $SU(4)$ singlet
state $|SGL \rangle$.  This $|SGL \rangle$ state contains 24 terms,
and is rotational invariant under the fifteen $SU(4)$
generators~\cite{YQLi98,Bossche00,Footnote1}. After tracing over the
orbital degrees of freedom, we find $S^{so}=1$ for this high
symmetry state. For the dimerized state at $(x=y=3/4)$, it is known
that its ground state is a matrix product state in both spin and
orbital part~\cite{Kolezhuk98}. After some algebra, we find its
value of entanglement is about 0.40.

\begin{figure}[t]
\includegraphics[width=8.0cm,height=7.2cm]{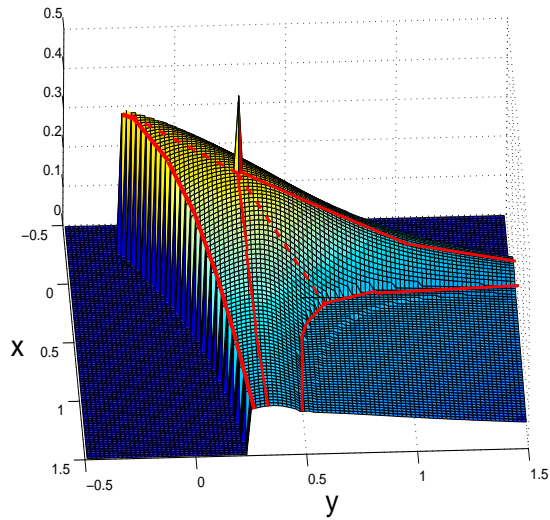}
\caption{\label{Fig1} The rescaled SOE $S^{so}/L$ $(L=8)$ as a
function of the $x$ and $y$. The phase boundaries (red lines) are
drawn to guide the eyes.}
\end{figure}

In what follows we will calculate the ground state SOE of the
Hamiltonian (1) in a finite size system.  We will demonstrate the
close connection between SOE and the quantum phase transitions in
the model. Since the Hamiltonian has a rotational symmetry around
the $z$-axes in $\mathbf{S}$-space as well as in $\mathbf{T}$-space,
the exact diagonalization calculations are carried out in an
invariant subspace with $S_z=0$ and $T_z=0$ to get the ground state
$|\Psi_G\rangle$, from which we construct the density matrix for the
whole system. The reduced density matrix $\rho_L$ of spin part is
obtained by tracing out the orbital degree of freedom, and compute
its reduced entropy. In our calculation, the chain length ranges
from 8 to 12.

The spatial profiles of SOE $S^{so}/L$ as a function of $x$ and $y$
are displayed in Fig. 1. A salient feature shows the existence of
zero entanglement regime. When either $\mathbf{S}_i$ or
$\mathbf{T}_i$ are aligned ferromagnetically, this model has no
frustration and qualitative results can be obtained from physical
considerations alone. Intuitively, for $x
> -1/4$ ($x < -1/4$) and $y < -1/4$ ($y> -1/4$), the ground states
are antiferromagnetic for spin $\mathbf{S}$ and ferromagnetic for
orbital $\mathbf{T}$ (and vice versa).  In the case of large
negative $x$ and $y$, the ground state is the ferromagnetic state
with respect to both $\mathbf{S}$ and $\mathbf{T}$. These three
conventional states correspond to the decoupling between the spin
and orbital degrees of freedom. Therefore, these states have zero
SOE. On the other hand, if both $\mathbf{S}$ and $\mathbf{T}$ are
coupled antiferromagnetically, the four-operator term may frustrate
the system, which may lead to the emergence of various non-trivial
ground states with finite values of SOE. As depicted in Fig. 1,
there are boundary lines with finite discontinuous jump of SOE
between the zero-value and finite-value regions which indicates a
first-order phase transition. Ground states of the boundary lines of
the fully ferromagnetic spin and orbital phases are highly
degenerate. It is obvious that the quantum fluctuation effect pushes
the classical phase boundary closer to the symmetric line $x=y$.

Two special points manifest themselves clearly: the $SU(4)$
symmetric point corresponds a local maximum of $S^{so}/L$ while the
dimerized state point corresponds a local minimum. This feature may
be simply understood as follows. At the $SU(4)$ point, there is the
largest correlation between spin and orbital degrees of freedom,
while at the dimer phase point, both the spin $\mathbf{S}$ and
orbital $\mathbf{T}$ are weakly coupled so that the entanglement is
much suppressed. In addition, the symmetric line $x=y$ is more
special and interesting. In Fig. 1, along the line (referred to as
the line $A$) connecting the point $(-1/4,-1/4)$ and the $SU(4)$
point where both of them have high symmetries, $S^{so}/L$ reaches
the local maxima ( ridge-like), while along the line (referred to as
the line $B$) connecting the $SU(4)$ point and (0.66,0.66), it
behaves as the local minima (valley-like). According to our previous
analysis and wisdom~\cite{chen04,chen05}, we know that both the
ridges and valleys may correspond to phase boundaries. We conclude
here that both lines $A$ and $B$ may serve as phase boundaries. It
is worth noting that the $SU(4)$ symmetric point is a multi-critical
point. There are four distinct neighboring quantum phases around
this point. For example, moving off this symmetric point, one may
enter the gapped states to the right upwards or enter the gapless
phases to the left downwards. These results are likely supported by
other studies. (i) The critical line A is consistent with that of
the analysis by Yamashita {\em et al.}~\cite{yam}. (ii) In a recent
Schwinger boson mean field study~\cite{Pengli05}, both spin and
orbital valence bond states are found in the parameter region
separated by the critical line B. Around the dimerized state point,
one may observe that there exists a curved phase boundary line
dividing the regions where the discontinuity of first derivative of
entanglement as a function of parameters occurs. The high
temperature series expansion approach suggests the existence of two
distinct gapped phases in the parameter region for both positive $x$
and $y$~\cite{Zheng01}. Our calculation supports the existence of
such gapped phases.

For the large $x$ and $y$ region, mean field studies always suggest
that the ground state is the antiferromagnetic state with respect to
$\mathbf{S}$ and $\mathbf{T}$. In that case, we would expect that
its corresponding SOE is equal to zero. However, the SOE in this
parameter region shows plateau-like behavior with finite value,
which contradicts the conclusion of mean field studies. Since the
well-known dimerized state point is located within the large $x$ and
$y$ region, we conclude that this phase regime belongs to the gapped
dimerized state rather than the gapless antiferromagnetic phase. It
is worth noting that the strength of SOE may be regarded as an
indicator to discern how good the mean field approximation will be.
In the strongly coupled regime, the interplay between spin and
orbital quantum fluctuations may be important and leads to some
highly nontrivial quantum phases. Therefore it is necessary to
consider the effects of quantum fluctuations more seriously beyond
the mean field theory. Certainly, we also find that both
entanglement measures vanish in the infinitely large limits of $x$
and $y$.

\begin{figure}[t]
\includegraphics[width=8.0cm,height=7.2cm]{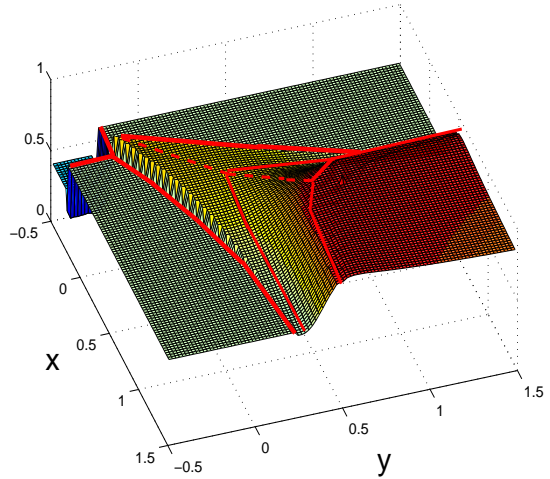}
\caption{\label{Fig2} The rescaled sublattice entanglement
$S_{L/2}/L$ $(L=8)$ as a function of the $x$ and $y$. The phase
boundaries (red lines) are essentially the same as that of Fig. 1.}
\end{figure}
To study  the additional entanglement between the intercalated
sublattices of composite degrees of freedom and to best reveal all
possible quantum phase boundaries, we also look into the standard
sublattice entanglement~\cite{chen04,chen05}, which is obtained by
tracing out both spin and orbital degrees of freedom at even (or
odd) sites in the present chain. In Fig. 2, we plot the sublattice
entanglement versus the coupling parameters. It is interesting to
note that there is roughly one-to-one correspondence of local
extreme and discontinuity between these two measures of
entanglement. In contrast to that of SOE, the $SU(4)$ point reaches
a local minimum of sublattice entanglement while the dimerized state
point corresponds to a local maximum. Since the SOE mainly captures
the correlation between the spin and orbital degrees of freedom
while the sublattice entanglement focuses on the correlation between
the intercalated sublattices of composite degrees of freedom, these
two measures may provide certain complementary information. In the
case of conventional ferromagnetic/antiferromagnetic phases, the SOE
vanishes while the sublattice entanglement remains nonzero. Thus the
measure of SOE is unlikely to distinguish these conventional phases.
Instead, in Fig. 2, these phases are clearly separated. In addition,
the enhancement of sublattice entanglement for the gapped dimerized
state is clearly observed.

\begin{figure}[t]
\hspace{5cm}
\includegraphics[width=7.6cm,height=6.6cm]{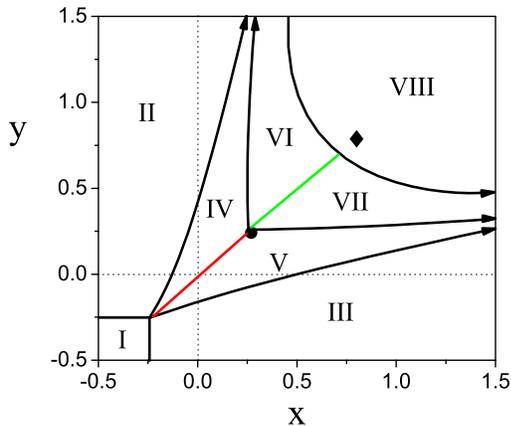}
\caption{\label{Fig3} Ground state phase diagram of a coupled
spin-orbital chain. The dotted point is at (1/4,1/4) while the
diamond point is located at (3/4,3/4). Eight distinct phases are
identified according to the analysis of entanglement as a function
of parameters $x$ and $y$. See text for details.}
\end{figure}

Quantum phase diagram can be distilled from the analysis of the
spatial profiles of entanglement as a function of parameters. In
other words, both the ridges and valleys in the three-dimensional
plot may correspond to possible phase boundaries. Derived from the
numerical results presented in Figs. 1 and 2, we plot the phase
boundaries of a coupled spin-orbital chain for $L=8$ in Fig. 3. The
results for $L=12$ are essentially the same. There are totally eight
distinct quantum phases. Most phase boundaries are in good agreement
with previous studies~\cite{Pati98,yam,Itoi00,Zheng01}. Phases I, II
and III, are conventional spin and orbital ferromagnetic or
antiferromagnetic states. Phase IV and V belong to gapless
states.~\cite{yam}. Phase VI and VII may correspond to orbital and
spin valence bond phases~\cite{Pengli05}, respectively. In view of
the fact that the exact ground state at the point $(3/4,3/4)$
belongs to staggered dimerized singlet and this dimerized state
point is located within phase VIII, we conclude that the phase VIII
is a staggered dimerized singlet state. It is remarkable that the
most comprehensive phase diagram is now efficiently and
straightforwardly obtained.

Since the coupled spin-orbital chain can be regarded equivalently to
a two-leg spin ladder with four-spin interactions, we may also
employ the measure of concurrence to quantify the bipartite
entanglement in terms of spin-spin, orbital-orbital as well as
spin-orbital concurrence. Our results show that the concurrence can
merely show a few features of phase diagram such as the conventional
phases I, II and III, but unfortunately, it fails to identify the
detailed phase diagram in the strong coupling regime. Another
scenario is to analyze  so-called single-site entanglement. In this
case, we obtain the reduced density matrix by tracing out all
degrees of freedom except for a single site and then get its reduced
entropy. However, we are still unable to identify many phase
boundaries. In our opinion, the failure of these two measures
highlights the importance of the nonlocal many body correlation
effect in characterizing some nontrivial quantum phases.

In conclusion, we present a novel approach to study phase diagram of
the coupled spin-orbital chain by coherently examining the
entanglement related to two distinctive degrees of freedom. The
analysis of the SOE supplemented by the sublattice entanglement
scenario enables us to establish an one-to-one link between its
local extreme/discontinuity and quantum transition points. The most
comprehensive phase diagram has been deduced for the first time
based on exact numerical results for a finite lattice system. Our
approach presents a superior and efficient way to identify quantum
phase transitions in a coupled spin-orbital system. The present work
may shed new light on the understanding of the complicated interplay
among charge, spin and
orbital degrees of freedom in transition-metal oxides 
in terms of entanglement.

The authors thank P. Zanardi, P. Li, S. Q. Shen and G. M. Zhang for
helpful discussions. This work was supported by the RGC grants of
Hong Kong, the RGC central allocation grant (HKU-3/05C), and Seed
Funding grants of HKU.


\begin{thebibliography}{}
\bibitem{Footnote1} The $SU(4)$ singlet state can be most conveniently written as, for the purpose
of studying SOE, $|SGL \rangle = \sqrt{2/3} [(12)_S (34)_S (14)_O
(23)_O -(14)_S (23)_S (12)_O (34)_O]$ where $(12)_{S(O)}$ stands for
the spin (orbital) singlet state of the sites 1 and 2.
\end{thebibliography}
\end{document}